\documentclass [12pt,twoside]{article}           % version LATEX2E

\usepackage{epsfig,times,lscape}
\usepackage[usenames]{color}

    \ifnum\count102=1

    \fi

    \ifnum\count102=2
\fi

\newcommand{\nc}{\newcommand}

     \ifnum\count101=1
\nc{\qI}[1]{\section{{#1}}}
\nc{\qA}[1]{\subsection{{#1}}}
\nc{\qun}[1]{\subsubsection{{#1}}}
\nc{\qa}[1]{\paragraph{{#1}}}

\def\qpar{\vskip 2mm plus 0.2mm minus 0.2mm}
\def\qL{\hfill \break}
     \fi 

      \ifnum\count101=2
 \nc{\qI}[1]{\parindent=0mm \vskip 8mm 
{\centerline{\LARGE \color{red}#1}}\vskip 3mm}
\nc{\qA}[1]{\vskip 2.5mm \noindent 
{{\bf\large\color{blue}  #1}} \vskip 1mm \parindent=0mm}
 \nc{\qun}[1]{\vskip 1mm \noindent {\sl #1 }\quad }

\def\qL{\hfill \break}
\def\qpar{\vskip 2mm plus 0.2mm minus 0.2mm}

      \fi
\def\qth{\vrule height 12pt depth 0pt width 0pt}
\def\qtb{\vrule height 0pt depth 5pt width 0pt}

\nc{\qfoot}[1]{\footnote{{#1}}}

      \ifnum\count101=1
\def\qbu{\hfill \par \hskip 6mm $ \bullet $ \hskip 2mm}
\def\qee#1{\hfill \par \hskip 6mm (#1) \hskip 2 mm}
      \fi
      \ifnum\count101=2
\def\qbu{\hfill \par \hskip 4mm $ \bullet $ \hskip 2mm}
\def\qee#1{\hfill \par \hskip 4mm (#1) \hskip 2 mm}
      \fi

\def\qparr{ \vskip 1.0mm plus 0.2mm minus 0.2mm \hangindent=10mm
\hangafter=1}

     \ifnum\count101=1 
 
     \fi
     \ifnum\count101=2

  \def\qcitb#1{\noindent \hbox to 102mm{\hfill \small #1} \vskip 1mm}
      \fi

 \def\qpages#1{\count102=0{\loop\advance\count102 by 1
 \null \vfill\eject \ifnum\count102<#1 \repeat}}

\def\qn#1{\eqno \hbox{(#1)}}

\def\qth{\vrule height 12pt depth 0pt width 0pt}
\def\qtb{\vrule height 0pt depth 5pt width 0pt}

\def\qv{\vskip 0.1mm plus 0.05mm minus 0.05mm}
\def\qhu{\hskip 0.6mm}
\def\qhv{\hskip 3mm}

\def\qhw{\hskip 1.5mm}
\def\qleg#1#2#3{\noindent {\bf \small #1\qhw}{\small #2\qhw}{\it \small #3}\qv }

\begin{document}
\thispagestyle{empty}
% --------------------------------------------------------------------

      % Hauts de pages et numerotation

          % Remarque: sans le \protect --> message d'erreur (ordre fragile)
\markboth{{\sl \hfill  \hfill \protect\phantom{3}}}
        {{\protect\phantom{3}\sl \hfill  \hfill}}

% -------------------------------------------------------------------
\color{yellow} 
\hrule height 20mm depth 10mm width 170mm 
\color{black}
\vskip -2.5cm 
\centerline{\bf \Large How are mortality rates affected by}
\vskip 2mm
\centerline{\bf \Large population density?}
\vskip 4mm

\centerline{\large 
Lei Wang$ ^1 $,
Yijuan Xu$ ^1 $,
Zengru Di$ ^2 $,
Bertrand M. Roehner$ ^{2,3} $
}

\vskip 15mm
\normalsize

{\bf Abstract}\quad Biologists have found that the 
death rate of cells in culture depends upon their
spatial density.  Permanent ``Stay alive''
signals from their neighbors seem to prevent them from dying.
In a previous paper (Wang et al. 2013)
we gave evidence for a density effect for ants.
In this paper we examine whether there is a similar effect
in human demography. We find that although there is no
observable relationship between population density 
and overall death rates,
there is a clear relationship between density and
the death rates of young age-groups. Basically their death rates
decrease with increasing
density. However, this relationship breaks down 
around 300 inhabitants per square kilometer.
Above this threshold the death rates remains fairly constant.
The same density effect is observed in Canada, France, Japan and the
United States. We also observe a striking parallel between
the density effect and the 
so-called marital status effect in the sense
that they both lead to higher suicide rates and are both 
enhanced for younger age-groups. However,
it should be noted that
the strength of the density effect is only a fraction of
the strength of the marital status effect.
In spite of the fact that this parallel does not give
us an explanation by itself, it
invites us to focus on explanations that apply to
both effects.
In this light the  ``Stay alive'' paradigm set forth by
Prof. Martin Raff appears as a natural interpretation. It 
can be seen as an extension of 
the ``social ties'' framework proposed at the end of
the 19th century by the sociologist Emile Durkheim
in his study about suicide.

\vskip 6mm
\centerline{\it First version: 20 May 2013. Comments are welcome}

\vskip 6mm
{\normalsize Key-words: Life expectancy, mortality, population 
density, suicide, group interaction, marital status.}
\vskip 2mm

PACS classification: Interdisciplinary (89.20.-a) +
collective effects (71.45.Gm)
\vskip 8mm

{\normalsize 
1: Laboratory of Insect Ecology, 
Red Imported Fire Ants Research Center, South China Agricultural
University, Guangzhou, China.\qL
Emails: wanglei\_1107@yahoo.com.cn, xuyijuan@scau.edu.cn
\qL
2: Department of Systems Science, Beijing Normal University, 
Beijing, China.\qL
Email: zdi@bnu.edu.cn 
\qL
3: Institute for Theoretical and High Energy Physics (LPTHE),
University Pierre and Marie Curie, Paris, France. \qL
Email: roehner@lpthe.jussieu.fr
}

\vfill\eject

\large

\qI{Overview}

\qA{The ``Stay alive'' paradigm}
It is not unreasonable to assume that for
any network of inter-connected living organisms
there is an ``optimum'' density of population. 
In the present study that ``optimum'' will be understood
in the fairly crude sense of allowing the longest life
expectancy. For {\it in vitro} cell populations
biologists have found that life expectancy decreases
strongly when the cell-density decreases%
\qfoot{Ishizaki et al. (1993 p. 904,  1994 p. 1072), Raff (1998).
See also the summary graph in Wang et al (2013, p. 5).}%
.
Prof. Martin Raff described this effect by saying that 
unless cells
permanently receive a ``Stay alive'' signal from their
neighbors, they are bound to die.
Is there a similar density effect in human populations?

\qA{Evidence at the level of US states}
At first sight the answer seems to be negative.
This is illustrated by the graph in Fig.1a which gives death
rates in fifty US states plus the District of Columbia.
The scatter plot does not show any significant trend.
As a matter of fact whatever correlation  might exist 
is due to the two points on the left- and right-hand sides,
namely Alaska and DC.
One may recall in this respect that the 
population of Alaska is markedly younger than
the average US population:
29.4 years as compared to 32.3 years for
the whole United States%
\qfoot{This example also serves to illustrate the 
well-known fact that one needs 
to examine the shape
of the scatter plot before drawing hasty conclusions, 
a methodology to which we will stick subsequently.}%
.
\qpar
  
Yet, if instead of total death rates one considers
the suicide rates of young people, one gets
a scatter plot which exhibits a highly significant
decrease with growing density, at least within the
range of densities covered in the graph of Fig. 1b.  
Just in order to get a more intuitive idea of what these
densities represent it can be mentioned that
10 inhabitant/sq.km corresponds to the case of
Nevada, 67 to Michigan, 94 to California, 330
to Massachusetts, 460 to New Jersey and 4,000 to
Washington DC. 
%
%%%% NUAGES DE POINTS: DENSITE - TAUX DE DECES
\begin{figure}[htb]
\centerline{\psfig{width=12cm,figure=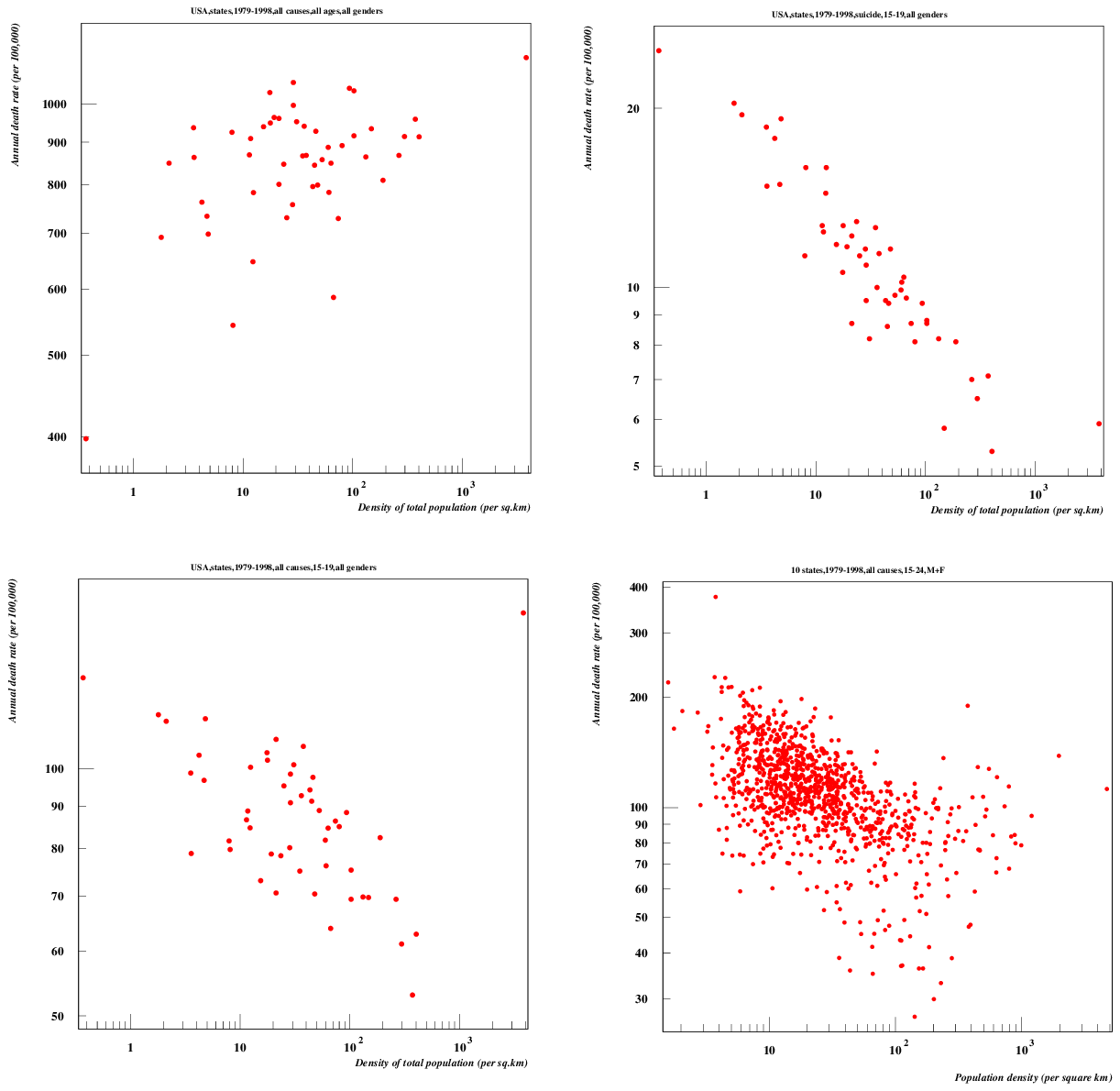}}
\qleg{Fig.\qhu 1a,b,c,d\qhv How does population density affect
death rates in the case of the United States, 1979-1998?}
{{\bf \color{blue} 1a}: All causes, all ages, US states (including DC);
 the (low) correlation 
is entirely due to two outliers, namely Alaska (lower left corner) 
and Washington DC (upper right corner).  
{\bf \color{blue} 1b}: Suicide, 15-19, US states; notice that the
horizontal scale
refers to the density of the {\it total} population, not the density
of the 15-19 age group; 
the correlation is $ -0.92 $ and the absolute value of
the slope of the log-log regression line is $ a=0.19\pm 0.02 $
(the error bar is for a confidence probability level of 0.95).
{\bf \color{blue} 1c}: All causes, 15-19, US states; 
the log-log regression gives
$ a= 0.052\pm 0.03 $.
{\bf \color{blue} 1d}: All causes, 15-24, 
1025 counties of 10 North-Central states 
(Georgia, Illinois, Indiana, Iowa, Kentucky, Michigan, Missouri, Ohio,
Tennessee, Wisconsin); the log-log regression gives
$ a=0.14\pm 0.01 $. The evidence from Fig. 1b,c,d
can be summarized by a law of the form $ r=r_0/d^a $ where 
$ r $ is the death rate and $ d $ the density. 
However, Fig. 1d suggests that after falling off sharply for
densities between 1/sq.km and $ d_c=300/ $sq.km, 
the death rate levels off for $ d>d_c $. A more precise
view can only be obtained by observing more high-density counties.
This will be done in Fig. 3.
All these results are for cumulative death numbers over the
20-year long time interval 1979-1998; similar results
are obtained for the periods 1968-1977 or 1999-2010.
For instance suicide in the 15-19 age group
leads to the following exponents: $ 1968-1977:\ a=0.15\pm 0.04,\quad
1999-2010:\ a=0.25\pm 0.02 $. Altogether over the
3 periods one gets the average exponent: $ a=0.20\pm 0.015 $.
Subsequently, for the sake of
simplicity and especially when they are under 10\% the error bars will
sometimes be omitted.} 
{Sources: Population and land area data: ``USA Counties'' data base
from the US Census Bureau; 
death rates: Centers for Disease Control and Prevention, National
Center for Health Statistics, Compressed Mortality File. }
\end{figure}
%----------------------------------------------

Does this mean that density has an influence on suicide only?
Not exactly. Fig. 1c shows 
a significant connection for ``all causes of death'' 
provided one focuses on 
young people. Yet, this relationship appears
weaker than the one for suicide.

\qA{Evidence at county level}

Many US states have large areas especially in the west.
A state like California comprises low density areas
as well as large urbanized areas around major cities.
The average population density cannot
be considered as a faithful indicator in such situations.
This is not the only drawback of an analysis 
that would limit itself to state-level data. 
Indeed, the graphs in Fig. 1a,b,c show that there are no
data points in the density interval between 460 (New Jersey) and
4,000 (Washington DC) inhabitants per square miles. In
other words state-level data do not allow us to
explore high density effects on death rates.
\qpar

These two reasons lead us to consider smaller spatial entities.
This was done in Fig. 1d which shows data for
some 1,000 of the 3,000 US counties. This graph broadly
confirms the effect suggested by Fig. 1c
\qfoot{Although the mean values of the exponents are not the same
their confidence intervals overlap and are not incompatible.}%
.
\qpar

Why did we draw this graph for North-Central states?
The reason is simple. These states hold
the largest number of counties and also the smallest counties.
Taken together Georgia, Indiana
and Kentucky have 366 counties whereas
California has 57 and Massachusetts only 6. The median land area
of Georgia's counties is 890 square kilometer against 6,200 for
those of the state of Utah. 
With smaller spatial entities the population density becomes a more
meaningful indicator. However, the other side of the coin is 
that together with smaller populations also come smaller numbers of deaths
which represents a serious
shortcoming for the investigation of specific causes of death%
\qfoot{For confidentiality reasons, the WONDER database omits
all counties for which there are less than 10 deaths. This
limitation represents a major obstacle for the analysis at county
level.}%
.
\qpar

The evidence presented in Fig. 1 speaks in favor of a relationship
of the form:
 $$ r={ r_0\over d^a },\quad 
\hbox{\small for } 1/\hbox{\small sq.km} \le d\le 300/\hbox{\small sq.km,}
\quad 
r:\hbox{\small  death rate,}\quad d:\hbox{\small  population density,} 
\qn{1} $$

Incidentally, it can be observed that defining $ a $ as a slope
in a {\it log-log plot} has the advantage of making it independent
of the measurement units used for the density or the death rate. 
\qpar

However,
several questions remain unanswered among which 
one can list the following:
\qbu Is there a gradual transition between the case of
young age groups for which there is a density effect and
the case of aged people for which none is expected?
\qbu Does the death rate curve level off above a density $ d_1 $
as Fig. 1d seems to suggest.
\qbu Is this effect specific to the United States or can
it be observed also in other countries?
\qbu Finally, one may wonder how this effect can be ``explained''.
\qpar

Before considering these questions more closely
one can give the following short answers.
\qbu From the 15-19 age group to the 74-84 age group
(the last group documented in the Wonder dataset)
there is a gradual decline in the exponent $ a $.
\qbu For densities above 300 per sq/km the death rate remains almost
constant.
\qbu Apart from the United States the 
density effect can also be observed in other countries.
We will show evidence for the cases of Canada, France and Japan.
\qbu A preliminary answer to the last question will be given
in the two sections before the conclusion section.

\qA{Death rate by cause as a function of age}

Before going closer into our investigation we wish to recall
how death rates increase along with age for different
causes of death. We will see that there is a sharp difference
between diseases and what the ``International Classification
of Diseases'' calls ``external causes of injury''.
Typically, as shown in Fig. 2,  
the death rates of diseases increase exponentially
with age (at least after the age of 30). On the contrary,
the rates for external causes hardly increase with age.
\qpar

Why is this distinction important for our investigation?
From Fig. 2 we see that the share of deaths
by disease or by external causes
is not at all the same in
young and in old age-groups. In young age-group deaths due
to external causes are predominant whereas deaths
due to diseases are predominant in old age groups.

%
%%%% TAUX DE DECES POUR 6 CAUSES EN FONCTION DE L'AGE
\begin{figure}[htb]
\centerline{\psfig{width=8cm,figure=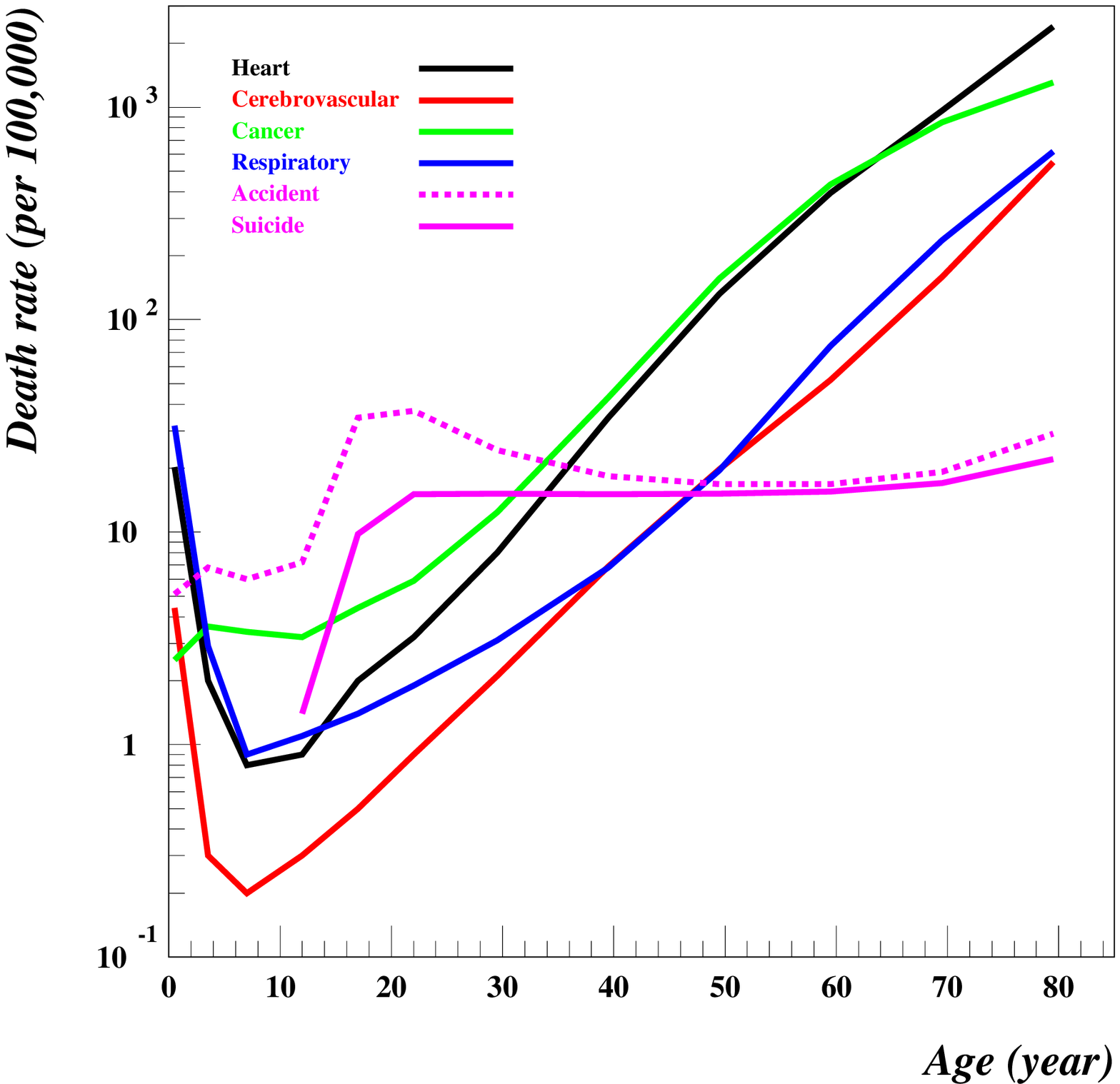}}
\qleg{Fig.\qhu 2\qhv Age-specific mortality rates for 6
causes of death in the United States, 1979-1998.}
{There are clearly two different groups: rates for cancer
and for heart, cerebrovascular, respiratory diseases increase
exponentially with age, whereas the rates for suicide and accident
remain constant or increase very slowly.}
{Source: Centers for Disease Control and Prevention, National
Center for Health Statistics, Compressed Mortality File. }
\end{figure}
%----------------------------------------------

Now, it turns out that in contrast to rates for external causes 
rates for diseases are almost {\it not}
affected by population density. Therefore one expects that
in an ``all causes'' analysis
young age-groups will be much more affected by density changes than
old age groups. This is indeed what was seen in Fig. 1.
In short, young versus old or external causes versus diseases
are just two aspects of the same effect.
\qpar 
However, it should be emphasized that this argument does not
tell us anything about how death from external causes are affected
by density or how the density effect itself is affected by age 
These points will be examined in the next sections.

\qI{Density effect for suicide}

By focusing on specific causes of death it becomes rapidly
apparent that the death rates for diseases are not affected
by density, or to say it more correctly,
there are other factors which have a stronger influence.
This can be seen fairly clearly by looking at the case of heart
diseases
at the level of US states plus Washington DC. If one discards
the two extreme cases of Alaska ($ d=0.5/ $sq.km) and 
DC ($ d=4,000/ $sq.km) there is no clear correlation among the
49 remaining states. However, the death rate in Alaska is
about one-half of the death rate in DC for young as well as
for old age-groups. This shows that some ethnic and/or
environmental factors are at work which are much stronger than
any possible influence of density.   
In other words if we wish to identify an effect of density
we must look either at causes of death that are not due to diseases
or at young age-groups for which these causes of death play
little role.
%
%%%% 4 GRAPHES EN FCT DE LA DENSITE POUR 4 CLASSES D'AGE
\begin{figure}[htb]
\centerline{\psfig{width=12cm,figure=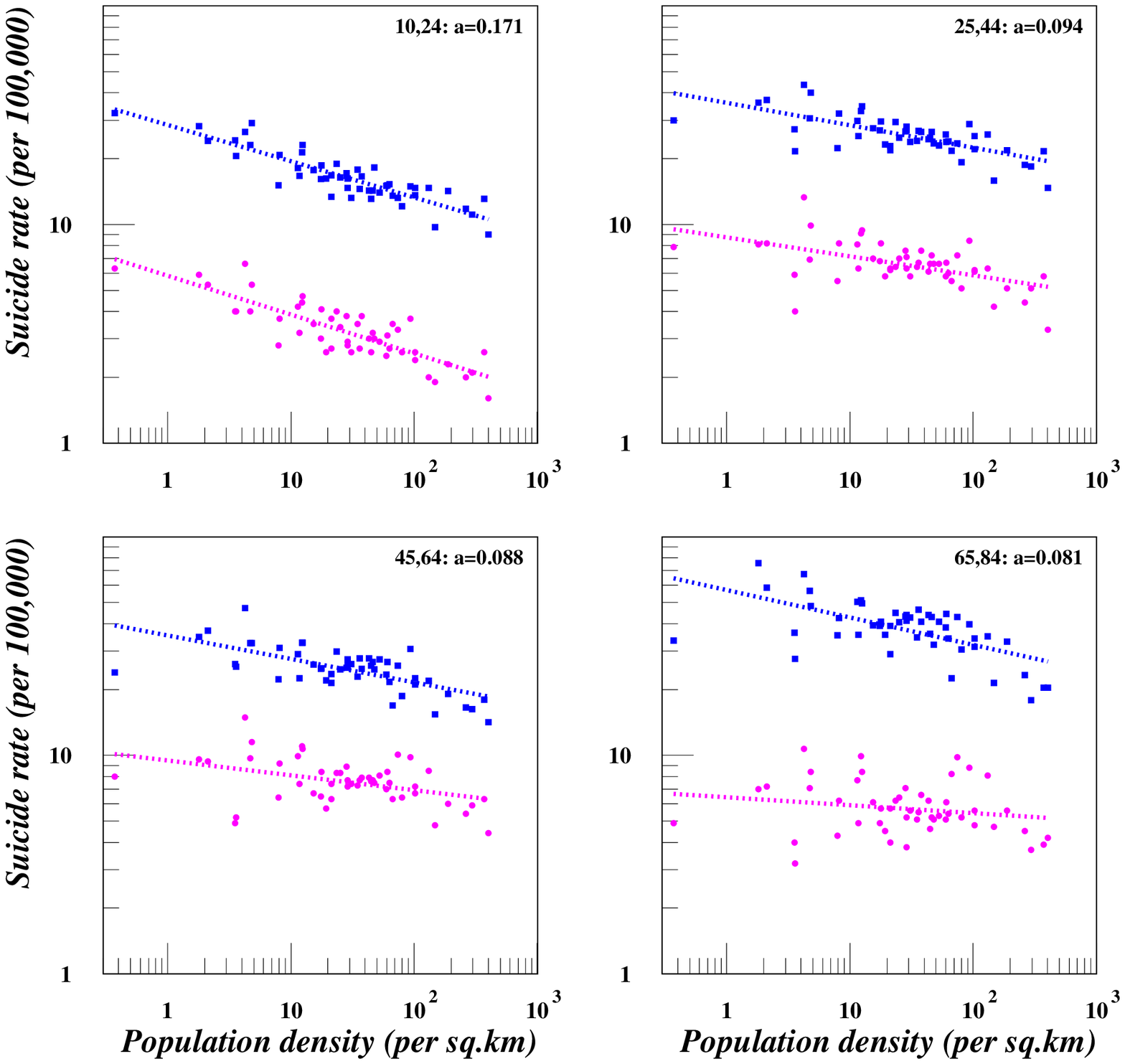}}
\qleg{Fig.\qhu 3\qhv Suicide rate in function of
density for 50 US states, 1979-1998.}
{Blue squares refer to males, magenta dots refer to females.
From young to old age-groups one observes a continuous decrease
in the slope $ a $ of the regression line. The numbers
in the upper-right corner are the averages of the male and female
slopes. At the same time the correlation becomes weaker: from
-0.89 to -0.62 for males and from -0.84 to -0.19 for females.}
{Source: Centers for Disease Control and Prevention, National
Center for Health Statistics, Compressed Mortality File. }
\end{figure}
%----------------------------------------------

There are mainly three causes of death due to external factors,
namely accidents, homicides and suicides. Which of them
is the most interesting for us? 
From the perspective of the apoptosis
paradigm introduced by Prof. Raff
\qfoot{In Raff (1998) he referred to
the phenomenon of cell apoptosis which happens either {\it in vitro}
or {\it in vivo} as being a kind of ``cell suicide''.}
it is clearly
suicide which will be of greatest interest.
However, accidents will also be discussed in some detail
later on.
\qpar

As far as suicide is concerned there are two
intriguing facts.
\qbu As already seen in Fig. 1 suicide rates are density-dependent
\qbu Secondly, the connection between suicide and
density becomes weaker and weaker as individuals become older.
\qpar
These two facts are documented in Fig. 3 in the case of
US states (this time without DC).  
The rates for males and females were plotted separately because
their orders of magnitude are fairly different. However,
their dependence with respect to density is
almost the same as shown by the fact that the regression lines
are nearly parallel. \qL
Weakening in the density-suicide interdependence can
best be measured by the decrease in the absolute value of the
correlation coefficient; the results are given
in the following table. 

%%-----------------------------------------------
\begin{table}[htb]

\centerline{\bf  Table 1:\quad  
Coefficients of correlation between suicide rate and population density}

\vskip 5mm
\hrule
\vskip 0.5mm
\hrule
\vskip 2mm

\color{black} 

$$ \matrix{
\qtb
\hbox{Age-group} \hfill & 10-24 & 25-44 & 45-64 & 65-84\cr
\noalign{\hrule}
\qth
\hbox{Male} \hfill & 0.89 & 0.70 & 0.67 & 0.62  \cr
\qtb
\hbox{Female} \hfill & 0.84 & 0.53 & 0.41 & 0.19  \cr
\noalign{\hrule}
} $$
\vskip 0.5mm
Notes: The table gives the absolute values of the coefficients
of correlation corresponding to the scatter-plots in Fig. 2.
The fall of the correlation for older age-groups occurs in
both genders but the reduction is much faster
for women than for men.
\vskip 2mm
\hrule
\vskip 0.7mm
\hrule
\end{table}
%%-----------------------------------------------

\qI{Evidence for low versus high density}

Fig. 1d has the advantage of being based on over 1,000 counties
but it has the disadvantage that these counties are mostly
in the (10,100) density interval. As a result, it does not
give good evidence for very low densities under 10 inh/sq.km
or very high densities over 300 inh/sq.km. In order
to explore these two ends, we built a special sample 
which includes
states with very low densities such as Arizona, Nevada, Utah or
Wyoming as well as states with very 
high densities such as Maryland, New Jersey, New York%
\qfoot{Altogether there are 13 states: Arizona, California,
Delaware, Massachusetts, Maryland, 
Michigan, Montana, Nevada, New Jersey, New York, Utah, Virginia,
Wyoming. These 13 states comprise only 447 counties
whereas the 10 states considered in Fig. 1d had twice as many.
It can be noted that
because of its large land area New York does not
have a very high density (only 160/sq.km) but it comprises several
top density counties.}%
. 

%
%%%% USA, SUICIDE, ALL AGES, DENSITE - TAUX DE DECES
\begin{figure}[htb]
\centerline{\psfig{width=10cm,figure=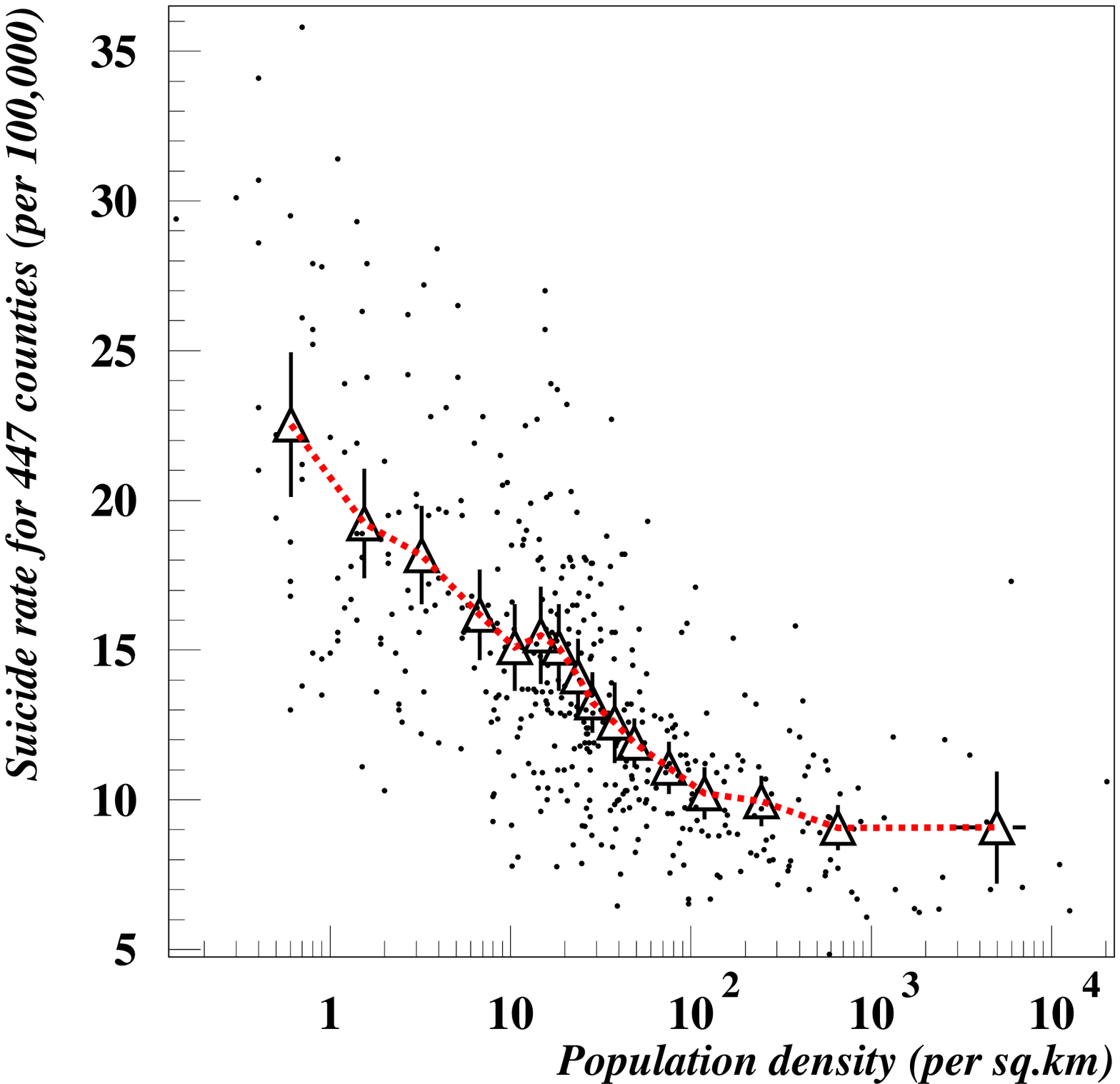}}
\vskip 2mm
\qleg{Fig.\qhu 4\qhv How population density affects 
suicide rates in the United States, 1989-1998.}
{The death rates are for suicide at all ages.
The broken line connects average for packages of
counties of increasing density.
We selected either low density states
such as Arizona or Utah or states with large
metropolitan areas such as New Jersey or New York
(the list of the states is given in the text).
As a result, one sees fairly clearly that the 
death rate curve has two distinct parts: a rapid
decrease (with a regression coefficient of -0.13)
under 200 inhabitants per sq.km followed by a plateau
for higher densities. } 
{Sources: Same as for Fig. 1}
\end{figure}
%----------------------------------------------

Fig. 4 shows that the density curve has two distinct parts:
one below 300/sq.km and a second one above this threshold
where the suicide rate remains fairly constant.
Fig. 5 and 7 given below show a similar effect for France and Japan.

\qI{Evidence for other countries than the United States} 

\qA{France}
For a country such as France whose population is about 5
times smaller than that of
the United States one faces the difficulty of a smaller number of
deaths. France is divided into 95 
{\it d\'epartements} and some 20,000 {\it communes}. Clearly,
at the level of the {\it communes}
the number of deaths would be far too small to be useful. Moreover, as
the number of {\it d\'epartements} is almost twice the number of US 
states, the average annual number of deaths per {\it d\'epartement}
will be some $ 5\times 2=10 $ times smaller than in the US.    
In other words, one is not in the best conditions to observe
a density effect. Nevertheless, a graph (Fig. 5) for death rates in
the age-group 15-24 gives a picture that is consistent
with Fig. 1c.
%
%%%% FRANCE, TOUTES CAUSES, 15-24, DENSITE - TAUX DE DECES
\begin{figure}[htb]
\centerline{\psfig{width=10cm,figure=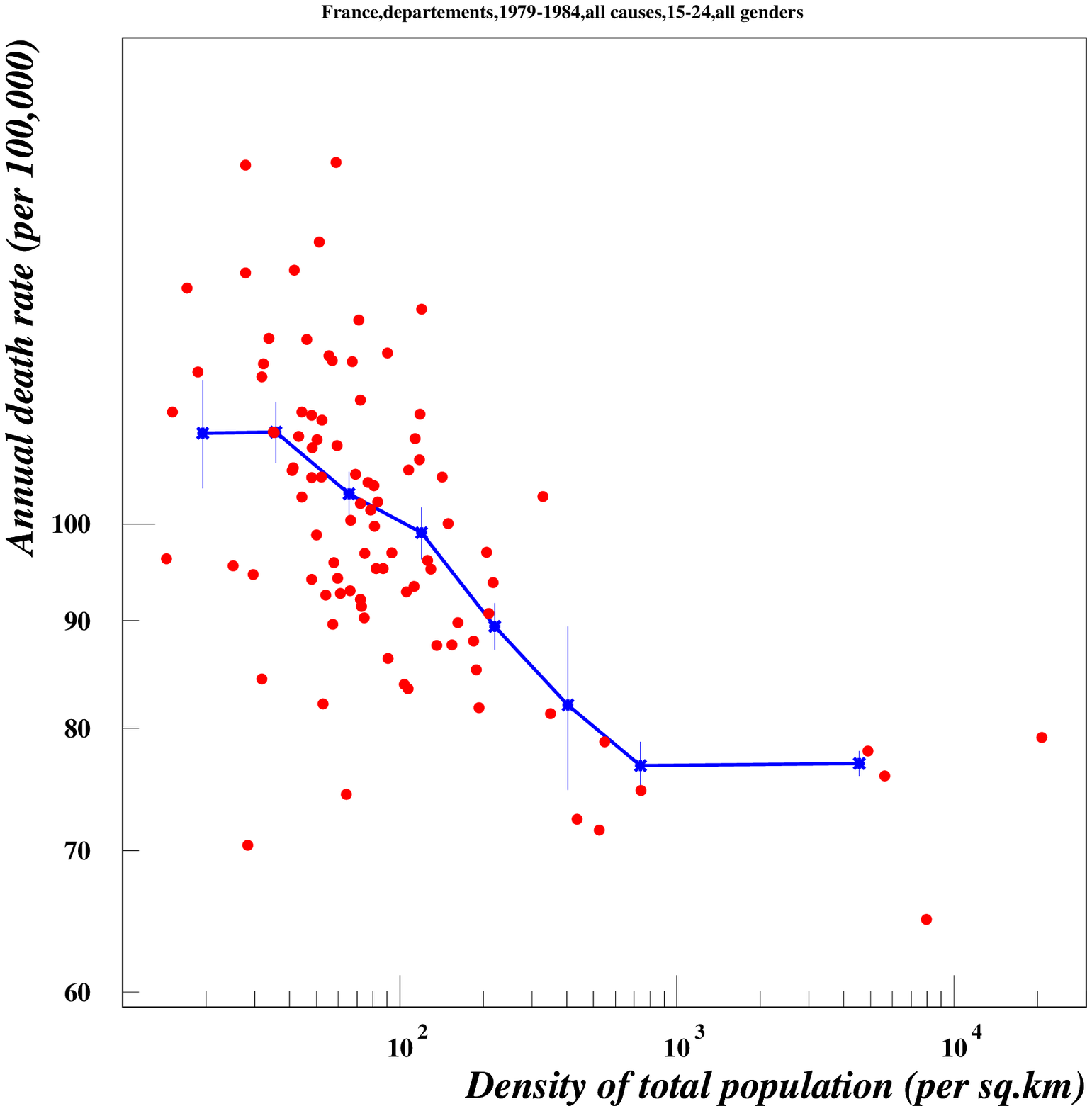}}
\qleg{Fig.\qhu 5\qhv How does population density affect 
death rates in the case of France, 1979-1984?}
{The death rates are for all causes of death in the age group 15-24. 
The slope of the regression line for the data points below
a density of $ 1,000 $ inhabitants per sq.km 
has an absolute value of 0.11. It appears that above this value
there is a leveling off but in order to get a more precise picture 
one would need more data points in this interval.} 
{Sources: Population and land area: INSEE; 
death rates by age: INSERM (``Institut national de la sant\'e
 et de la recherche m\'edicale'' which is the French analogue of the
US ``National Center for Health Statistics'').}
\end{figure}
%----------------------------------------------
\qA{Canada}

In the late 20th century Canada had a population of some
30 millions and its territory is divided into 12 provinces
and territories. 

%
%%%% CANADA, DENSITE - TAUX DE SUICIDE
\begin{figure}[htb]
\centerline{\psfig{width=8cm,figure=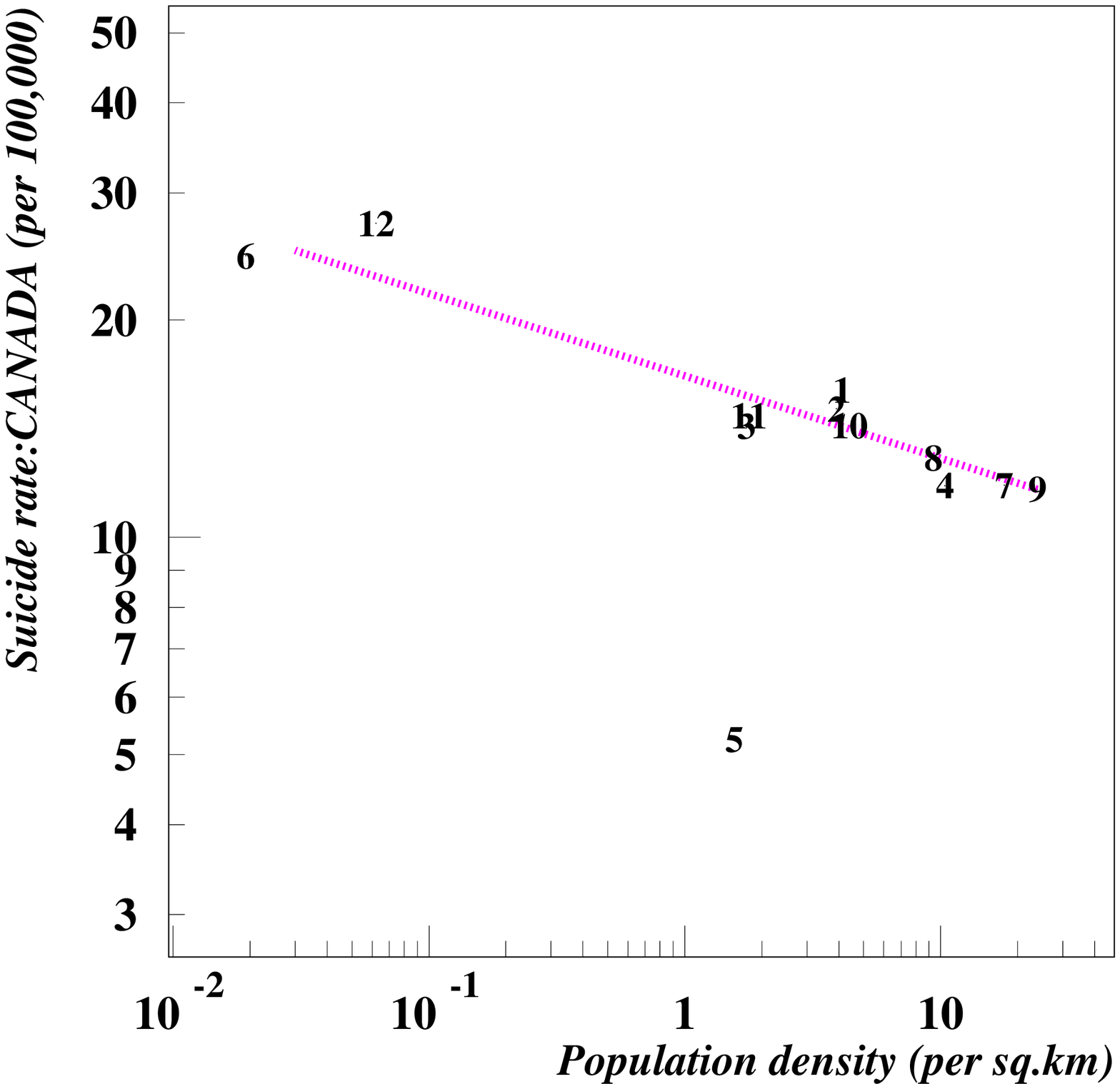}}
\qleg{Fig.\qhu 6\qhv How population density affects 
suicide rates in the case of Canada, 1971-1990.}
{Whereas the case of France gave us a glimpse of the
incidence of high densities, Canada gives us information
about the low density range . The graph suggests
that the power law found in previous cases remains valid
for the very low densities found in North-West (no 6)
and Yukon (no 12) Territories. Newfoundland (no 5) appears to
be an obvious outlier (its suicide rates are closer to
those which prevail in the UK than to those
seen in the rest of Canada). With Newfoundland left apart,
the log-log regression gives an exponent $ a=0.11\pm 0.05 $
which is consistent with values found previously.
The large error bar does not come from a poor correlation
(it is equal to $ -0.83 $) but from the small number of
data points.}
{Source: Health Canada (1994).}
\end{figure}
%----------------------------------------------

Compared with the United States the expected annual
number of deaths per province will be 
$ (300/30)\times (12/50)\simeq 2 $.
This makes the situation less unfavorable than in the case
of France but at the cost of having but a tiny number of data
points. Moreover, all these regions have population densities
which are low or very low. There are in fact two distinct groups:
(i) a group of 10 provinces whose densities are approximately
comprised between
1 and 10 inh/sq.km (ii) a group of two territories, namely
the North-West Territories and Yukon, whose densities are 
below 0.1 inh/sq.km.
A report of Health Canada (1994)
gives suicide data over a 30-year period from 1961 to 1990.
\qpar
In spite of the few data points the graph in Fig. 6 covers
a broad range of population densities. Thanks to the fact
that it is an average over a 20-year long period, its
suicide rates can be considered as fairly reliable.
If one leaves aside Newfoundland, an obvious outlier for
some unknown reason, one gets an exponent $ a=0.11 $.  
\qpar

As a function of age-group one gets the following results%
\qfoot{As the death rates of the N-W and Yukon Territories
become somewhat unstable in separate age-groups  
(due to small numbers), they were left aside.}%
:
$$ 15-24: a=0.20\pm 0.07 \quad 
35-54: a=0.067\pm 0.09 \quad 65-74: a=0.12\pm 0.13  $$

Of these three age-groups, it is the first one which has the
most significant connection with population density.

\qA{The case of Germany}
With a  population of 80 millions in 2010, Germany may appear
as a good candidate for the present investigation. In fact,
it is not. There are three main reasons for that.
\qbu The United States has an average population density of 28/sq.km
(in 1996) whereas Germany has an average density of 234/sq.km (in
2010). Thus, it is not surprising that
of the 16 German {\it L\"ander} none has a density under 70/sq.km.  
As seen above 
the decrease of death rates as a function of density occurs
mainly between 0.1/sq.km and 30/sq.km. Over 70/sq.km one would
expect only a small residual decrease.
\qbu The {\it L\"ander} which have the lowest densities% 
\qfoot{Brandenburg (85), Mecklenburg-Vorpommern (86), Saxony-Anhalt
  (119), Thuringia (143).}%
are all former East German {\it L\"ander}. For that reason 
their suicide rates are not really comparable to those in
former West German {\it L\"ander}.
\qbu The time series of death rates in former East German {\it
 L\"ander} start after the reunification in 1990. Thus, the largest
interval over which one can perform averages runs from 1990 to
present time.
\qpar

If, despite such low expectations, one nevertheless draws the curve
of average suicide rates over 1990-1997
as a function of density one gets the following
results. Though seemingly disappointing, they are in fact consistent with
our expectations.
\qee{1} For all ages the slope is slightly (non significantly)
negative: $ a=0.012\pm 0.09 $ (the correlation is -0.07).
\qee{2} For the age-group 15-19 the correlation is slightly
higher, namely -0.25 (yet still not significant) and 
$ a=0.040\pm 0.08 $.
\qpar

Needless to say, even if the correlation had been somewhat higher
the fact that there are only 16 {\it L\"ander}
would still result in giving to $ a $ a large confidence interval.

\qA{The case of Japan}

For Japan the story is very much the same as for Germany
in the sense that all 47 prefectures have a population density
higher than 70/sq.km. Hokkaido has the lowest (72 in 2005) and there are
only two others with a density under 100, namely Iwate (91) and Akita
(99). Therefore, if the law $ r=f(d) $ 
is indeed the same as seen above one
expects only a small residual decrease of the suicide rate between
the densities of Hokkaido at one end and a threshold 
density $ d _c $ which should be around 300/sq.km.
\qpar
What makes the situation markedly better than for Germany
is the fact that we have here 47 prefectures instead of
only 16 {\it L\"ander} 
\qpar
Actual observation for suicide rates at all ages 
averaged over the time interval 2009-2011 
shows results that are indeed consistent with expectation.\qL
There are two distinct parts: a downward trend under $ d_1=300 $
and an horizontal line above this threshold.
For the 26 prefectures whose density is under $ d_c $ the
correlation is $ -0.56 $ with a confidence interval
$ (-0.78,-0.22) $ which shows that it is significantly
negative (despite the fact that there are only 26 data points).
The log-log regression estimate of the exponent
is $ a=0.19\pm 0.11 $.

\qpar 
It would have been interesting to 
see if the density effect is amplified by
restricting the age to the 15-19 age group. So far, however,
we were not able to get the required statistics.  

%
%%%% JAPON, DENSITE - TAUX DE SUICIDE
\begin{figure}[htb]
\centerline{\psfig{width=8cm,figure=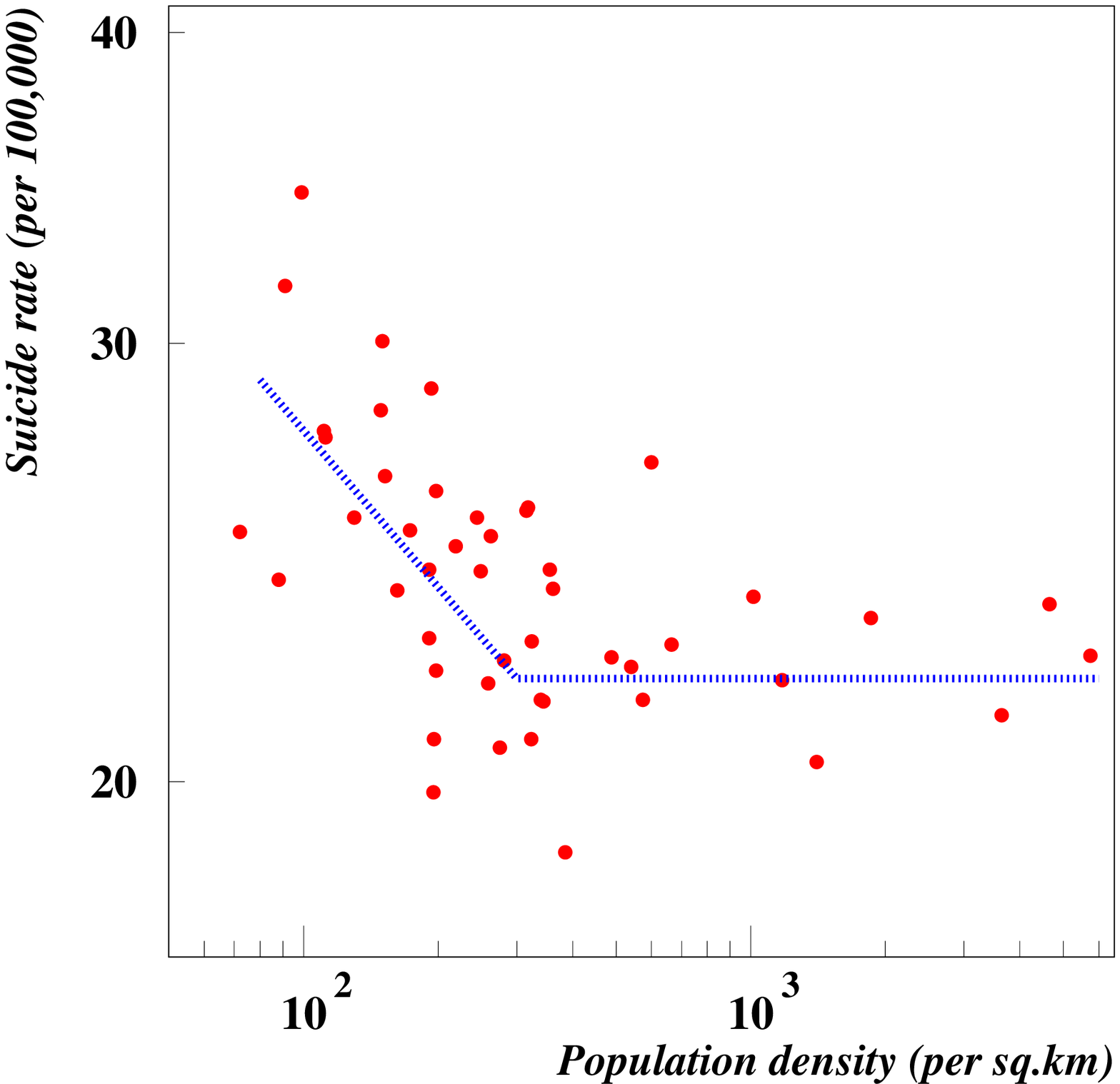}}
\qleg{Fig.\qhu 7\qhv How population density affects 
suicide rates in the case of Japan, 2009-2011.}
{Because Japan's 47 prefectures have mainly high or very high
densities this graph shows a limited section of the downward curve
together with a long level line corresponding to 
densities over a threshold $ d_c=300/ $sq.km. 
For the 26 prefectures which have
a density under $ d_c $ regression analysis gives
an exponent $ a=0.19\pm 0.11 $.}
{Source: Population densities: 
website entitled ``Historical Statistics of
Japan''; suicide rates: website entitled ``Vital Statistics''
of the Ministry of Health, Labour and Welfare.}
\end{figure}
%----------------------------------------------

\qI{Determinants of the density effect}

\qA{What kind of explanation?}

There are (at least) three possible kinds of explanations for the
density effect described in the previous sections.
\qee{1} It may be a statistical artifact. 
\qee{2} It may be explained by some fairly obvious anthropomorphic 
factors.
\qee{3} The effect may have a deeper (non anthropomorphic)
origin, for instance in relation with the ``Stay alive''
effect mentioned at the beginning of the paper.
\qpar
Let us examine these possibilities more closely.

\qA{Statistical artifact}
The best guarantee against possible
statistical  artifacts is to repeat the observation in different
countries in the hope that they do not all use the same
statistical methodology.
So far, we have considered four countries but in fact we gave
a look at quite a few others. These attempts
convinced us that is is not easy to find many ``good''
candidates. This can be illustrated by the following discussion
about Sweden as well as by the case of Germany that we 
already considered.
\qpar

Sweden is a country which is often useful for statistical 
investigations because of its excellent statistical system.
In 2010 Sweden had a population of some 9 million
and a territory divided into 284 municipalities.
This means that compared with the United States
the annual number of deaths per municipality
will be smaller by a factor
$ (300/9\times (284/50)=189 $. In other words,
in order to get the same cumulative number of deaths one
would have to consider a time-period that is 189 times
longer than in the US. As in the US we have used time-periods
of at least 10 years this is clearly impossible.
\qpar

The previous argument suggests that one should rather turn to
countries with large populations. Among them China appears
as an obvious choice. That sets a possible objective for
a subsequent investigation.   
\qpar

\qA{Anthropomorphic factors}
One obvious factor
comes to mind immediately. For persons who have an accident or a 
heart attack it will take longer to take them to an hospital
if they live in a fairly desert country side than if they
live in a city. However, that effect should affect old people
as well as young people. It could even be argued that the effect
should be stronger for old people because they are less
resilient and would therefore be more affected by a long
delay to get to the hospital. Yet, for old people
one does not observe any clear density effect. In short,
there does not seem to be a ``distance to hospital'' effect.
\qpar

For deaths through accidents and especially car accidents 
there is undoubtedly a density effect. In cities the average 
velocity of cars is low which means that collisions will rarely 
result in fatal injuries. On the contrary, on countryside roads
the high speed reached by cars will transform 
any collision into a fatal accident. This intuitive argument
is indeed confirmed by statistical observation. In all European
countries the graph of traffic fatality rates as a function of 
population density shows the same downward trend (Orselli 2001).
For young drivers this countryside road effect may be amplified
due to poorly developed driving skills. 
\qpar

Needless to say, the previous argument applies only to road traffic
accidents and cannot explain the downward trend of suicide deaths.
However, as traffic accidents are the first cause of death in the
15-24 age group, this factor certainly plays a role
in this age range, albeit a limited role because 
in past decades this cause of death 
represented only between 27\% and 35\% of all deaths in this
age group. Another point which remains unclear 
is the fact that one does not know
which fraction of the downward trend is due to the velocity factor 
{\it per se} as compared with other density-dependent factors.

\qI{The ``Stay alive'' effect}

The ``Stay alive'' mechanism easily ``explains'' the fall
of suicide rates when the density increases. However,
it does {\it not} explain why the fall stops around $ d_c=300/ $sq.km;
neither does it explain
why young age-groups are more affected than older
age groups.
\qpar
Let us consider these points more closely. One important
aspect that will be developed is the parallel between the
effects of population density and those of marital status.

\qA{Less links means more suicide}

We mentioned at the beginning that the death rate of cells
in culture increases when their density becomes lower. The results
given in  Ishizaki (1992, 1993) suggest that the death rate is
multiplied by 2 or 3 (depending upon cell type)
when the density is divided by 10. In Fig. 1 we have seen that 
the death rate of young people is multiplied by a factor of about 2
when the density is divided by 100.
In any population the frequency of contacts 
between individuals is proportional to the square 
of their density. Thus, smaller density means less contacts%
\qfoot{Of course, smaller density may also have other effects,
but the reduction of contact frequency seems to be a property that
is shared by many systems.}%
.
\qpar

At this point we need to explain why, among various causes of death,
suicide plays a special role.
It is well known that among the different causes of death it is 
suicide which is the most sensitive to the lack or severance 
of social links. This can be seen by looking at how death rates depend
on marital status. Whereas for heart disease or cancer the
death rate of bachelors or widowers 
is on average some 1.8 times larger
than for married people, in the case of suicide the average 
excess-death ratios
jump to 2.5 for bachelors and 5 for widowers%
\qfoot{This can be seen on Fig. 3b of Wang et al. (2013).}%
. 

\qA{Amplification of the density effect for young age-groups}
For different marital situations 
age-specific suicide ratios tell us something about
the role of age.

%
%%%% GRAPHE EN FONCTION DE L'AGE, // AVEC MARITAL STATUS
\begin{figure}[htb]
\centerline{\psfig{width=8cm,figure=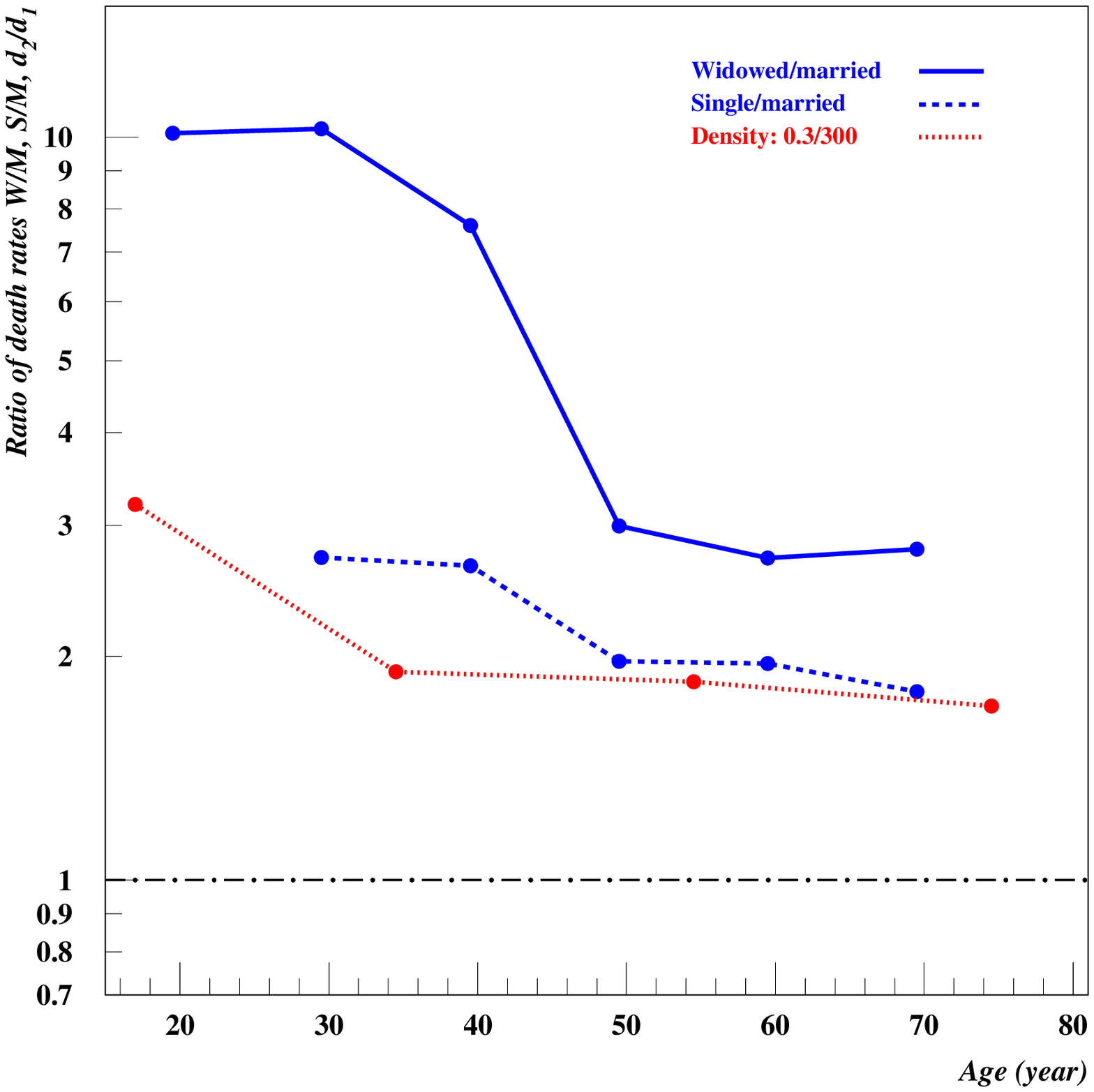}}
\qleg{Fig.\qhu 8a\qhv Parallel between the effect on
suicide rates of marital status on the one hand and
of density level on the other hand.} 
{The marital status ratios are for the United States in 1980.
In the density case
the ratios were derived from the linear regression
performed in Fig. 3; the ratio is the suicide rate for a density
of 0.3/sq.km divided by the suicide rate for a density of 300/sq.km.
The latter has been selected because it corresponds to the
minimum suicide rate in the same way as the suicide rate
of married people is the minimum among family situations.
What is the practical meaning of the 0.3/300 curve? 
It means that if 1,000
persons of same age (e.g. 40) 
move from a place $ A $ where the density is 300/sq.km
to a place $ B $ where the density is 0.3/sq.m, they will
experience $ r $ times (for age 40, $ r=2 $) more suicides than
a control group of same age that has remained in $ A $.
All data in this graph are for males and females together.}
{Sources: Suicide rates by marital status: Vital
Statistics of the United States, 1980, Vol. 2, part A, p. 323;
Suicide rates by age and population density:
Centers for Disease Control and Prevention, National
Center for Health Statistics, Compressed Mortality File.}
\end{figure}
%----------------------------------------------

%
%%%% GRAPHE EN FONCTION DE L'AGE, TOUS LES DECES
\begin{figure}[htb]
\centerline{\psfig{width=10cm,figure=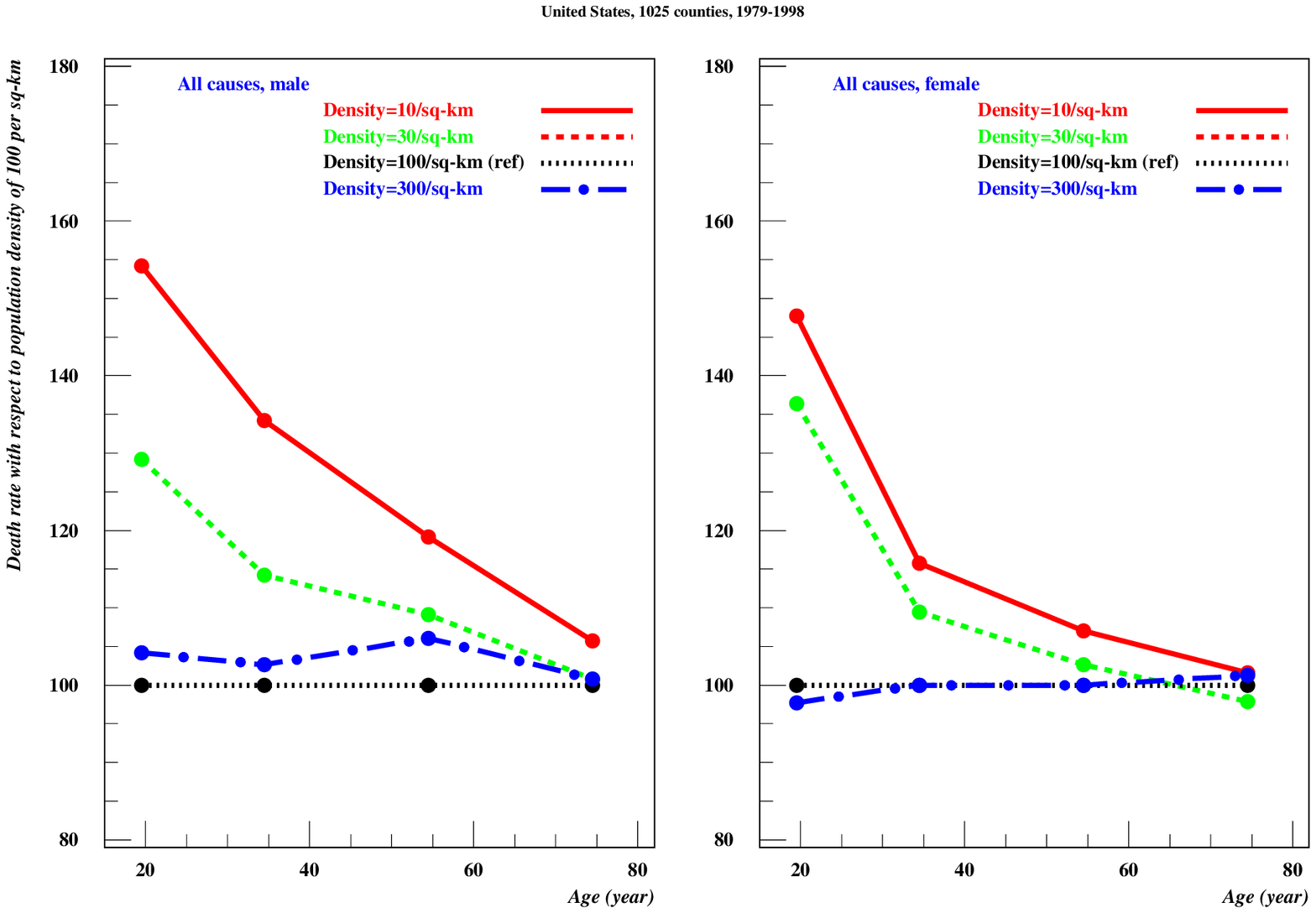}}
\qleg{Fig.\qhu 8b\qhv Effect on
death rates of density level and age} 
{This graph is similar to the density curve of Fig. 8a
but instead of suicide rates it considers death rates
from all causes. It is based on an analysis at county level. 
The different curves give
the ratios: density $ d $/density 100 for $ d=10,\ 30,\ 300 $;
the index was normalized to 100 for a ratio equal to 1. 
It can be seen that
the results are almost the same for a density of 100 and 
for a density of 300. 
In contrast with Fig. 8a, 
for age-groups over 60 the ratio tends toward 1 which
means that for these age-groups there is no density effect.
This is due to the fact that
old-age deaths are mostly deaths from diseases.}
{Sources: 
Centers for Disease Control and Prevention, National
Center for Health Statistics, Compressed Mortality File.}
\end{figure}
%----------------------------------------------

They show an amplification of the effect of
tie severance for {\it young}
widows or widowers% 
\qfoot{There is no similar pattern in the
case of heart disease or cancer; on the contrary, the excess
mortality is reduced for young individuals.}%
. 
This pattern is illustrated in Fig. 8a and
Fig. 8b. For Fig. 8a the
suicide rates were derived from the regressions 
on US states performed in Fig. 3. 
If 
one denotes the regression line by $ \ln s=a_i\ln d+b_i $
where the index $ i $ refers to the 4 age-groups, then
the suicides rates for the lowest and highest density
(along with their ratio) will be given by: 
$$ S_i=\exp \left(a_i \ln(0.3)+b_i\right),\quad
s_i=\exp \left(a_i \ln(300)+b_i\right),\quad r_i={ S_i\over s_i } 
\quad i=1,\ldots ,4 $$ 

For Fig. 8b, the death rates were derived directly from
an analysis done at county level for the same 1,025 counties 
already used  in Fig 1d. 
This was made possible  
because the number of deaths is much larger than
the number of suicides. For suicides as a function of age
the analysis is almost impossible at county level just
because they are too few.
\qpar 

Of course, by itself the parallel drawn between density and marital
status does not offer an explanation. However, it suggests
that any explanation that would not apply to both cases may
not be satisfactory. In other words, it narrows the range of
possible explanations. 
\qpar

Regarding the amplification for young age-groups,
we can offer an additional observation.
Basically, the argument goes as follows. As
suicide rates of young age-groups are 
driven up by the effect of low density, 
one would expect that the 
curves of rates as a function of age are {\it flatter} in places
of low density than in places of high density.

%
%%%% GRAPHE EN FCT DE L'AGE, SECOND // AVEC MARITAL STATUS
\begin{figure}[htb]
\centerline{\psfig{width=11cm,figure=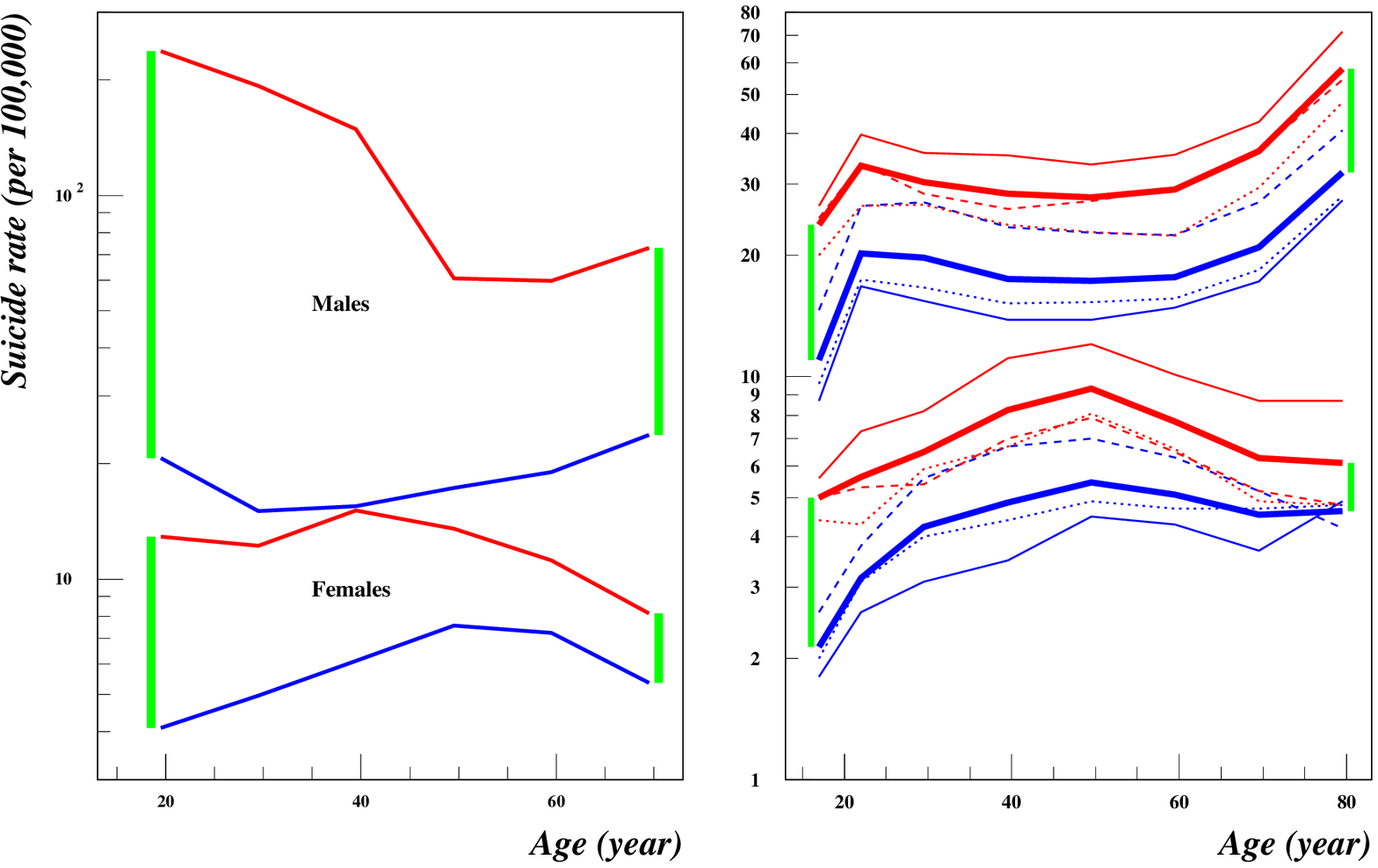}}
\qleg{Fig.\qhu 9\qhv Graph showing the way in which lower density
(or severed marital ties) affect age-curves.} 
{The curves for males are on top of the curves for females.
The red curves show cases characterized by reduced links:
widows and widowers on the left, low density areas on the right.
The blue curves show networks characterized by higher 
interactions: married individuals on the left, 
high density areas on the
right. For the density case, it would have been fairly
arbitrary to select just one state of each kind. Instead
we have selected 3 areas of each kind. The low
density areas are the following. 
A (solid line)=(AZ,CO,NE,NV,NM,WY),
B (broken line)=(ID,MT,ND,SD), C (dotted line)=KS. 
The high density areas consist in
the following states: A (solid line)=NJ, B (broken line)=PA,
C (dotted line)=NY. The thick solid lines are averages
of the A,B,C cases. Overall, the curves for low density
areas are flatter than those for high density areas.
As a result, the ratios at 20 year of age are larger than
the ratios at 80. These ratios are represented by the green
vertical bars. On the marital status graph 
the same effect can be observed but in much stronger form. 
The upper curves which correspond to widowers and widows
are not only flatter, they are in fact inverted in the sense
that instead of going up with age they fall off.    
Basically, this figure is  equivalent
to Fig. 8. but it is seen from a different, more basic, angle.}
{Sources: Suicide rates by marital status: Vital
Statistics of the United States, 1980, Vol. 2, part A, p. 323;
Suicide rates by age and population density:
Centers for Disease Control and Prevention, National
Center for Health Statistics, Compressed Mortality File.}
\end{figure}
%---------------------------------------------- 
\qpar
In order to make this point clearer and to see if it is really true,
we have drawn a graph (Fig. 9). It shows that age-curves
in low density areas are indeed flatter than in high density places.
These graphs, once again, underline
the parallel with marital status.
\qpar

In the following section we tentatively propose
a mechanism which accounts for what one observes in Fig. 9.

\qA{An explanation in terms of shock-strength}

How can one explain that the suicide rate of young widowers
is two or three times higher than the suicide rate of
widowers over 50? Was the shock represented by the death of
their spouse more painful for young widowers than for
widowers over 50? At first sight one would rather expect
the opposite. Indeed, it can be argued that a marriage that has
lasted for 20 or 30 years creates a stronger link than a
marriage that has been in existence for just a few years.
This is merely wishful thinking however. A more solid
argument is to estimate the propensity for getting married
by the marriage rate (defined as the number of annual
marriages in a given age group by the number of non-married
persons in the same age group)  
just in the same way as one would measure
the rate of a chemical reaction. This
calculation shows that the marriage rate is
highest at age 28 and decreases sharply thereafter. At age 60
it is about 10 times lower than at 28 (Roehner 2008, p. 72-74).
If one accepts that the propensity for getting married provides
a measure of the strength of the bond%
\qfoot{This is the way of thinking used in chemistry where
the strength of a bond is estimated through the
energy required to create it or to break it.}
,
then one comes to a conclusion which is opposite to
the previous one. In this perspective the high suicide
rate of young widowers is in line with the strength of the
shock that they experience.
\qpar

Is it possible to transpose this argument to the density case?
The analog of the widowhood shock would be the transfer from
an urban environment that corresponds to a density of over
500/sq.km to a countryside environment. How can one
estimate the strength of the links that connect 
residents to their social environment? Obviously, one
cannot use the same method as for marriage. Another
indicator that can be used is the suicide rate itself. 
For a given density individuals in old age-groups have
a higher suicide rate than young people. 
If one agrees with the perspective set forth by
Emile Durkheim (1888, 1897), this suggests 
that their connection with their familial and social environment
is weaker than that of younger people%
\qfoot{Intuitively, this seems of course very plausible because
of the strong link that an occupational activity 
represents for individuals until
they retire. Incidentally, Fig. 9 shows that the suicide rate
of men begins to climb up after the age of 60 whereas for
women the curve remains fairly flat.}%
.
Now comes the last step in our argument.
If young persons are more connected to their social
environment, then the shock of moving out to a countryside
environment will be more painful to them and, just as in the case
of widowhood but with smaller magnitude, it will result
in inflated suicide rates.
\qpar

A word of caution is perhaps in order with respect
to the previous argument.
Once one has identified a possible mechanism (here the
parallel with marital status), it is
generally difficult to prove that it is indeed the right one.
In this respect, our proposal
should be seen rather as a working hypothesis. If, in the course
of time, it appears that it is able to explain 
an ever growing set of observations (rather than just some isolated facts)
this will make it more convincing and better accepted.

\qI{Conclusions}
This paper gave statistical evidence for the following effects.
\qee{1} Whereas there is usually no clear connection
between {\it overall} death rates and population density ($ d $),
a significant relationship turns up between $ d $ and the
death rates ($ r $) in young age groups.
\qee{2} Whereas overall death rates are dominated by the deaths
of people over 60, the death rates for suicide or for accidents
are rather dominated by the deaths of young or middle-aged people%
\qfoot{This is true even when the suicide rate increases
with age because the age-groups over 60 represent a 
smaller share of the total population than the groups
of young or middle-aged people.}% 
. 
Therefore it is not surprising that suicide or accident rates
exhibit a strong connection with density even if all
ages are included.
\qee{3} In those cases (defined above) where the death rate 
falls off with increasing density, the decrease slows down
for densities over 300 inh/sq.km and is followed by a plateau.
\qee{4} The pattern described by the previous rules is
observed in very similar ways in Canada, France 
Japan and the United
States. It can be summarized by a power-law of the form:
$ r=r_0/d^a $ where $ a $ is of the order of 0.12 for suicide
rates and somewhat lower for all-causes death rates among
young individuals. 
\qee{5} The ``Stay alive'' paradigm% 
\qfoot{It can be seen as an extension of the framework
set forth by Emile Durkheim (1888, 1897). That point of view
was also developed in Roehner (2007, part 3)}
establishes a connection
between death rates and strength of inter-individual interactions.
Fig. 8 and 9 establish a parallel between the effects of
marital status on the one hand and population 
density on the other hand which suggests that the strength
of intra-family contacts is similar to the strength of
social (non-familial) interactions.
\qpar

At this point we do not wish to claim that this
pattern holds in a general way and in all times.
Whereas family links have remained fairly unchanged
in recent time, big changes have affected social life 
in villages and towns of industrialized countries.
Back in the 19th century, countryside villages and
small towns were still vibrant places of living characterized
by a broad spectrum of activities, from farmers to craftsmen or
clerks. With the advent of very large mechanized farms,
the social network of the countryside has lost much of its diversity
and interactions. In the United States this 
transformation mostly occurred during
the first half of the 20th century. In western Europe it occurred
mainly during the second half of the 20th century.
In many developing countries, and in particular in China,
it is currently occurring at great speed under our eyes.
\qpar
 
The purpose of the present article was not to explain everything
but rather to describe the empirical pattern and to  
state as clearly as possible the interrogations
which remain. Once we get a
clearer understanding of the factors which control this effect,
we will be able to predict the situations in which it should
occur as well as those in which it will not be expected.

\vskip 7mm
{\bf References}

\qparr
Centers for Disease Control and Prevention, National
Center for Health Statistics, W
Compressed Mortality File. [commonly referred
to as the ``WONDER data\-ba\-se''.]

\qparr
Durkheim (E.) 1888: Suicide et natalit\'e. Etude de statistique
morale. Revue philosophiques de la France et de l'\'etranger,
vol. 26, 446-463. [An English translation was published
under the title: ``Suicide and the birth rate. A study in
moral statistics'' (1996), Barclay D. Johnson, Ottawa.]

\qparr
Durkheim (E.) 1897: Le suicide. Etude de sociologie. F.Alcan, Paris
[A recent English translation is: ``On Suicide'' (2006),
Penguin Books, London.]

\qparr
Health Canada 1994: Suicide in Canada. Update of the report of
the task force on suicide in Canada. Published by the Minister of
National Health and Welfare. [The data are contained in
Appendix 6: section 1 (number of suicides) and section 2 (suicide
rates, (p. 119-202).]

\qparr
Institut National de la Sant\'e et de la Recherche
M\'edicale [French National Institute for Health and
Medical Research]: Database for statistics on causes
of death [partly available on line].

\qparr
Ishizaki (Y.), Voyvodic (J.T.), Burne (J.F.), Raff (M.C.) 1993:
Control of lens epithelial cell survival.
The Journal of Cell Biology 121,4,899-908 (May)

\qparr
Ishizaki (Y.), Burne (J.F.), Raff (M.C.) 1994: Autocrine
signals help chondrocytes to survive in culture.
The Journal of Cell Biology 126,4,1069-1077 (August)

\qparr
 Ministry of Health, Labour and Welfare of Japan: Vital
Statistics of Japan. [available on line]

\qparr
Orselli (J.) 2001: Les Fran\c{c}ais ne sont pas si mauvais
conducteurs. La Recherche 342, May 2001, 68.
[The paper discusses the influence of regional population
density on the number of victims of car accidents in various
European countries.]

\qparr
Raff (M.C.) 1998: Cell suicide for beginners. Nature 396, 12 November,
119-122.

\qparr 
Roehner (B.M.) 2007: Driving forces in physical, biological
and socio-economic phenomena. A network science investigation
of social bonds and interactions. Cambridge University Press,
Cambridge.

\qparr
Roehner (B.M.) 2008: Interaction maximization as an evolution
principle for social systems. Lectures given in China
and Japan in September-December 2008. LPTHE working paper,
University Pierre and Marie Curie, Paris. [also
available at the library of Sciences Po.] 

\qparr
US Census Bureau: USA Counties. [available on line at the following 
address:\qL
http://censtats.census.gov/usa/usa.shtml]

\qparr
Wang (L.), Xu (Y.), Di (Z.), Roehner (B.M.) 2013: How does group
interaction affect life expectancy?
Preprint, arXiv (Los Alamos) reference number 1304.2935.

\end{document}